\documentclass[a4paper,fleqn]{cas-dc}

\usepackage[numbers]{natbib}

\def\tsc#1{\csdef{#1}{\textsc{\lowercase{#1}}\xspace}}
\tsc{WGM}
\tsc{QE}
\tsc{EP}
\tsc{PMS}
\tsc{BEC}
\tsc{DE}

\begin{document}
\let\WriteBookmarks\relax
\def\floatpagepagefraction{1}
\def\textpagefraction{.001}
\shorttitle{CMOS BEOL compatible FTJ devices}
\shortauthors{V. Deshpande et~al.}

\title [mode = title]{CMOS back-end-of-line compatible ferroelectric tunnel junction devices}                      



\author[1]{Veeresh Deshpande}



\author[1,2]{Keerthana Shajil Nair}
\author[1,2]{Marco Holzer}
\author[1]{Sourish Banerjee}




\author%
[1,2]
{Catherine Dubourdieu}

\address[1]{Helmholtz-Zentrum Berlin für Materialien und Energie, Hahn-Meitner-Platz 1, 14109 Berlin, Germany}
\address[2]{Freie Universität Berlin, Physical Chemistry, Arnimallee 22, 14195 Berlin, Germany}



\begin{abstract}
Ferroelectric tunnel junction devices based on ferroelectric thin films of solid solutions of hafnium dioxide can enable CMOS integration of ultra-low power ferroelectric devices with potential for memory and emerging computing schemes such as in-memory computing and neuromorphic applications. In this work, we present ferroelectric tunnel junctions based on Hf$_{0.5}$Zr$_{0.5}$O$_{2}$ with materials and processes compatible with CMOS back-end-of-line integration. We show a device architecture based on W-Hf$_{0.5}$Zr$_{0.5}$O$_{2}$-Al$_{2}$O$_{3}$-TiN stacks featuring low temperature annealing at 400°C with performance comparable to those obtained with higher temperature annealing conditions. 

\end{abstract}
%



\maketitle

\section{Introduction}

The recent developments in solid solutions of hafnium dioxide ferroelectric materials have enabled integration of ferroelectric materials in CMOS technology. Ferroelectric materials have the potential for development of ultra-low power consumption and fast switching memory devices with excellent retention properties. Among different device architectures, ferroelectric tunnel junctions (FTJs) and ferroelectric field-effect transistors (FeFETs) allow non-destructive read-out and scalability to nanoscale dimensions. This has led to a strong interest in the development of CMOS integrated ferroelectric devices with HfO$_{2}$-based ferroelectrics. There has been significant development in FTJ and FeFET technologies recently~\cite{1}. However, as FeFET is essentially a front-end-of-line device that shares silicon area with other CMOS circuits, the potential for ultra-dense embedded memory integration on CMOS is limited. FTJs on the other hand are simple two-terminal devices, that can be integrated into the back-end-of-line (BEOL) of CMOS technology. This would enable vertical stacking of memory layers on top of CMOS circuits. BEOL integration dictates the thermal budget to be limited to around 400-450°C. Therefore, the crystallization temperature needed to form the ferroelectric phase in HfO$_{2}$-based films needs to be limited to this range. 

Hf$_{0.5}$Zr$_{0.5}$O$_{2}$ (HZO), with its low crystallization temperature compared to other HfO$_{2}$ solid solutions, becomes the primary choice for BEOL integration. While there have been recent reports of FTJ devices with a HZO ferroelectric layer~\cite{2,3}, the thermal budget utilized is typically 500-600°C. In this work, we demonstrate FTJ devices featuring 400°C crystallized HZO films with tungsten (W) as one of the electrodes. Tungsten electrode based FTJs can be seamlessly integrated into CMOS BEOL at W contact plug level or middle-of-line metallization level. We report on the electrical behavior of the FTJs with electrical cycling and under various pulse width programing conditions.

\section{Device design and fabrication}

\begin{figure}
	\centering
	\includegraphics[scale=.6]{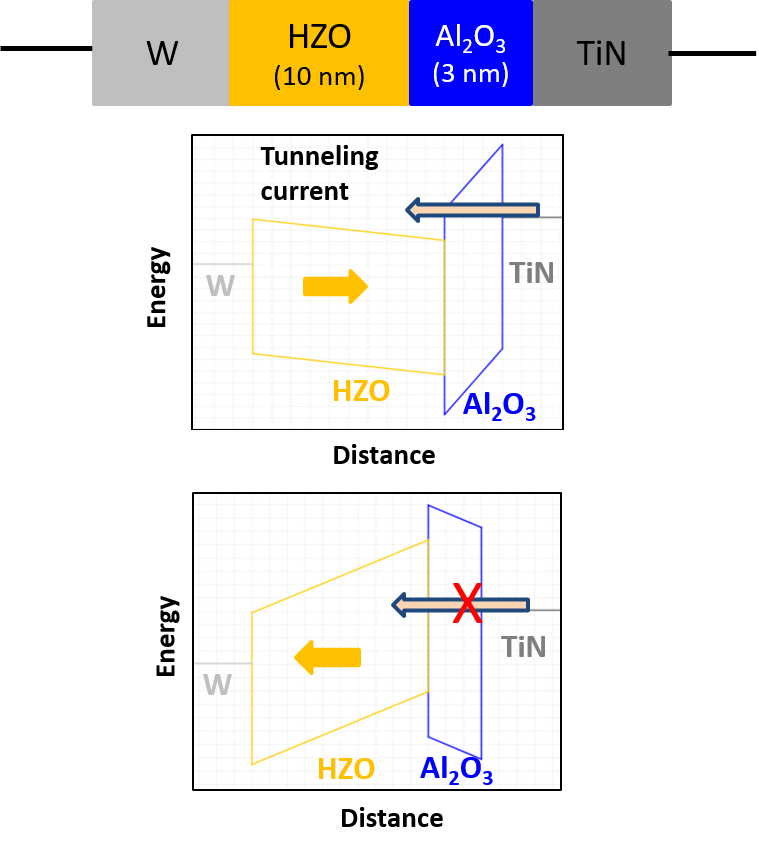}
	\caption{Band diagrams showing band bending for two polarization directions for W-HZO-Al$_{2}$O$_{3}$-TiN stack.}
	\label{FIG:1}
\end{figure}

The simplest FTJ device stack typically comprises of a Metal-Ferroelectric-Metal stack with dissimilar metals on either side of the ferroelectric layer. The tunneling rate across the ferroelectric layer is different for the two polarization directions thereby giving different tunneling currents (or resistances) for the two polarization states. The tunnel electro-resistance (TER) ratio quantifies the relative difference between the two currents. As the main phenomena governing device operation is tunneling, the thickness of the ferroelectric layer is limited to 1-3~nm to obtain reasonable tunneling current. However, obtaining a high remnant polarization (necessary for reasonable TER) in such ultra-thin HZO layers is quite challenging. Only recently, there have been research works to demonstrate ferroelectric behavior in such ultra-thin films~\cite{4}. One can also obtain reasonable TER with relatively thick HZO layer (8-10~nm) by utilizing an alternate device architecture consisting of Metal-Ferroelectric-Dielectric-Metal stack. The dielectric layer has to be ultra-thin to allow tunneling current across it. The band diagrams for the two polarization directions are shown in the schematic in Fig.~\ref{FIG:1}. When the polarization is pointing towards the dielectric layer (Al$_{2}$O$_{3}$ here), the voltage drop inside the dielectric reduces the barrier width and allows electrons to tunnel through to the conduction band of the ferroelectric layer. When the polarization is pointing away from the dielectric layer, the potential drop in the dielectric and ferroelectric layer is such that the effective barrier width is large enough to block the tunneling current. Therefore, one can achieve a large difference between the tunneling currents for the two polarization directions.
The FTJ stack presented in this work is W-HZO-Al$_{2}$O$_{3}$-TiN. The top electrode is W. The devices are fabricated on top of a p++ doped silicon substrate and all the measurements are carried out with bottom contact on the substrate. First, a 30~nm TiN layer is deposited by sputtering (at room temperature). Then the Al$_{2}$O$_{3}$ layer (3~nm) and the HZO layer (10~nm) are deposited by atomic layer deposition at 250°C. Finally, the 30~nm top W electrode layer is deposited by sputtering at room temperature. The whole stack is annealed by RTP at 400°C for 120s in N$_{2}$ ambient. This serves as crystallization anneal for the HZO layer. Square pads of 95 x 95~µm$^{2}$ area are patterned on the top W layer by photolithography. In an alternate process, called ‘metal replacement’ process, a stack of TiN-HZO-Al$_{2}$O$_{3}$-TiN is first deposited. The stack is annealed in the same RTP conditions as mentioned before. After the annealing, the top TiN is etched in SC1 solution and a 30~nm W layer is sputtered. Square pads (of 95 x 95~µm$^{2}$ area) are subsequently patterned.

\section{Results and Discussion}
In order to verify the ferroelectric properties of the HZO layer, a TiN (30~nm)-HZO (10~nm)-TiN (30~nm) stack was first characterized using the same RTP conditions. The HZO layer needs about 2000 cycles for wake up. A remnant polarization (measured with PUND pulse sequence) of about 15~µC/cm$^{2}$ is obtained after wake-up. This is similar to many reports for 10 nm-thick HZO films~\cite{5}.

\begin{figure}
	\centering
	\includegraphics[scale=.75]{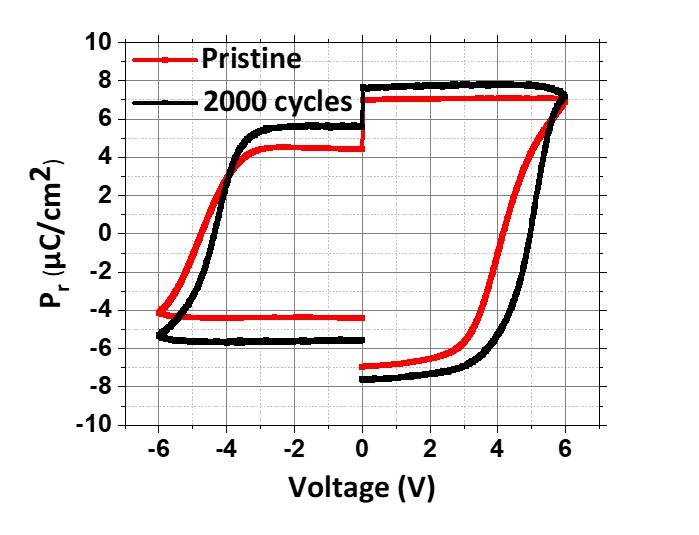}
	\caption{Polarization-Voltage (P-V) curves for W (30~nm)-HZO (10~nm)-Al$_{2}$O$_{3}$ (3~nm)-TiN (30~nm) stack for pristine state and after 2000 cycles wake up (PUND measurements).}
	\label{FIG:2}
\end{figure}

The P-V characteristics (measured with PUND sequence) of a FTJ stack with W (30~nm)-HZO (10~nm)-Al$_{2}$O$_{3}$ (3~nm)-TiN (30~nm), annealed with W top electrode are shown in Fig.~\ref{FIG:2}. As in the TiN-HZO-TiN capacitor case, the stack shows wake-up behavior. A remnant polarization of about 8~µC/cm$^{2}$ is reached after 2000 cycles, which is much lower than that of the TiN-HZO-TiN stack. This is due to the fact that the dielectric layer can only partially compensate, at its interface with HZO, the bound charges associated to the polarization. The offset in polarization at V = 0 V is due to the asymmetry of the interface on either side of HZO layer, leading to different polarization magnitude for the two polarization directions.
 
\begin{figure}[t]
	\centering
	\includegraphics[scale=.5]{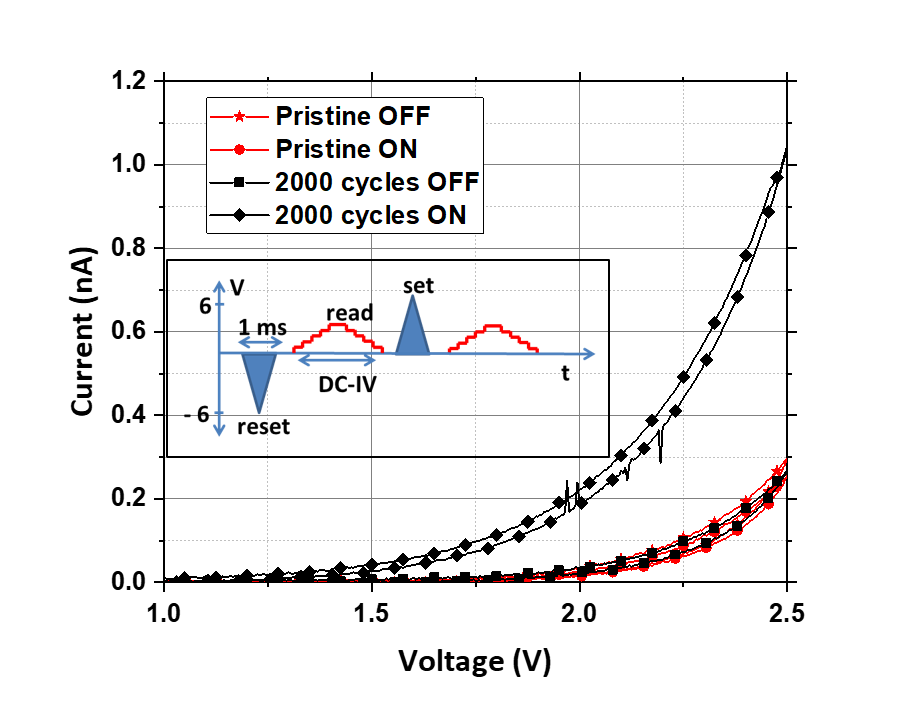}
	\caption{Current-Voltage (I-V) DC measurements of ON and OFF read currents at pristine state and after 2000 cycles for W (30~nm)-HZO (10~nm)-Al$_{2}$O$_{3}$ (3~nm)-TiN (30~nm) stack. Inset shows the pulse sequence.}
	\label{FIG:3}
\end{figure}

Figure~\ref{FIG:3} shows the representative current-voltage (I-V) characteristics of the FTJ stack after ‘set’ and ‘reset’ pulses. The reset (-6~V) and set (6~V) operations are performed with a triangular pulse of 1 ms width. The ‘read’ I-V operation is a DC measurement where the voltage is swept quasi-statically to 2.5~V. The sequence of measurement is shown in the inset. After 2000 cycles wake-up, there is a clear difference between the ON and OFF state currents. The resultant TER is about 3. The ON state current is ~0.01~mA/cm$^{2}$ at 2.5~V, which is similar to recent reports with a higher annealing temperature~\cite{2}.

\begin{figure}[b!]
	\centering
	\includegraphics[scale=.75]{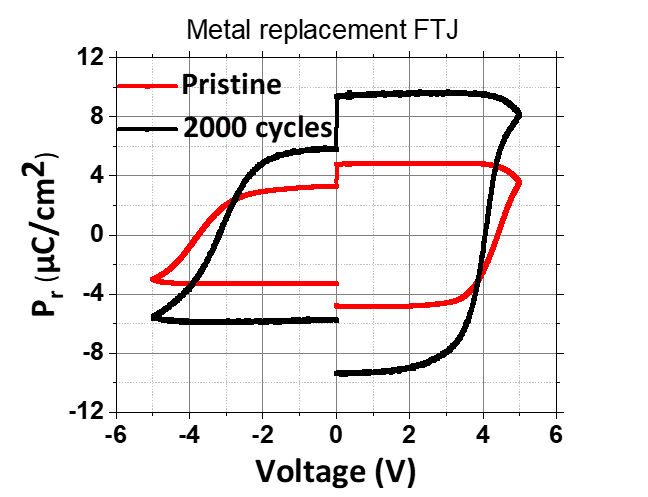}
	\caption{Polarization-Voltage (P-V) curves (PUND measurements) for W (30~nm)-HZO (10~nm)-Al$_{2}$O$_{3}$ (3~nm)-TiN (30~nm) ‘metal-replacement’ stack for pristine state and after 2000 cycles wake up.}
	\label{FIG:4}
\end{figure}

\begin{figure}
	\centering
	\includegraphics[scale=.55]{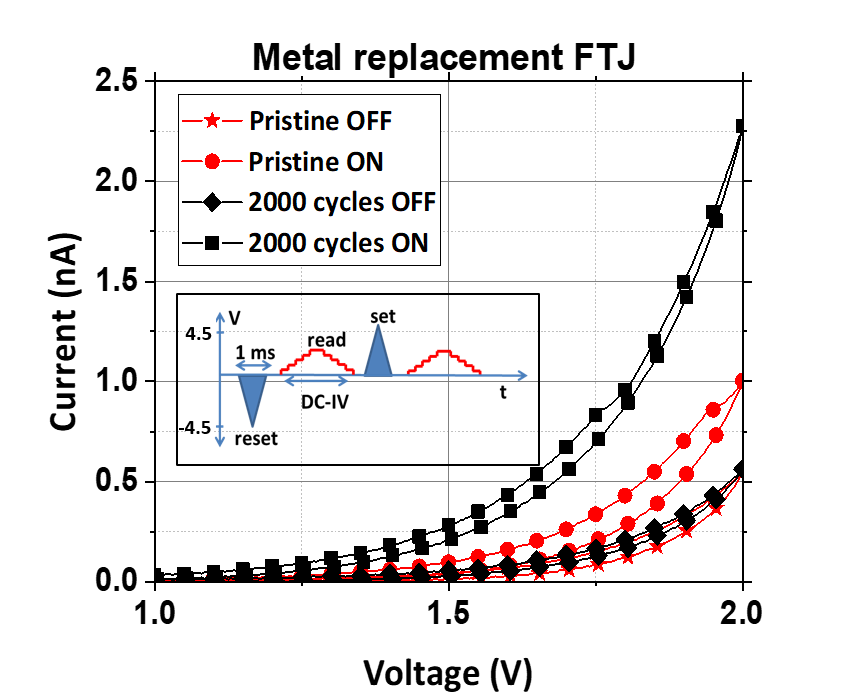}
	\caption{Current-Voltage (I-V) DC measurements of ON and OFF read currents at pristine state and after 2000 cycles for W (30 nm)-HZO (10 nm)-Al$_{2}$O$_{3}$ (3 nm)-TiN (30 nm) ‘metal-replacement’ stack. Inset shows the pulse sequence.}
	\label{FIG:5}
\end{figure}

As mentioned in the previous section, another FTJ stack was fabricated with ‘metal replacement’ process. In this stack, annealing was done with TiN as top metal first and was subsequently replaced by W. The P-V characteristics (measured with PUND sequence) of the stack are shown in Fig.~\ref{FIG:4}. As observed in the FTJ stack annealed with a W top electrode, there is a wake-up phase observed after 2000 cycles. The I-V measurement of the tunneling current for ON and OFF states is shown in Fig.~\ref{FIG:5}. The set/reset pulse sequence is the same as the one used for the previous stack (shown in the inset schematic). Note that the voltage magnitude necessary to fully switch the polarization in this stack is $\pm$~4.5~V with coercive voltage in the range of $\pm$~3 to 4.5~V. Therefore, the read voltage is limited to 2~V. In ‘metal replacement’ FTJ stacks, the ON and OFF state currents are different even in the pristine state. The resultant TER is about 1. After 2000 cycles, the TER increases to ~3.5 and the ON current density is about 0.024 mA/cm$^{2}$ at 2~V, twice as high as that measured in FTJ stacks annealed with a W top metal.

\begin{figure}
	\centering
	\includegraphics[scale=.55]{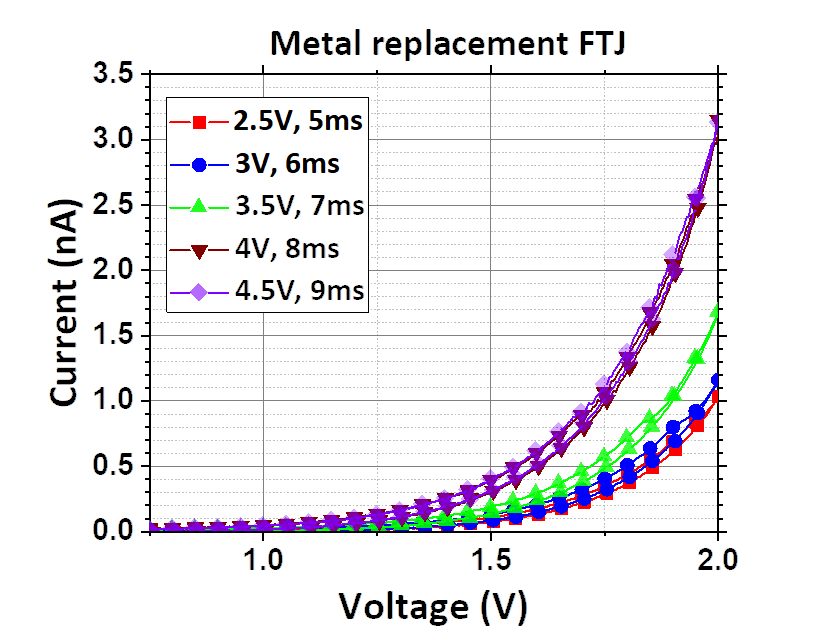}
	\caption{Intermediate ON-resistance state characterization. The set operation is done with different voltages and the pulse widths are adjusted to keep the same sweep rate.}
	\label{FIG:6}
\end{figure}

Emerging computing applications such as neuromorphic computing and in-memory computing with novel memory devices require multiple memory states in one device. In case of FTJs, it can be achieved by partial polarization switching due to e.g. a distribution in coercive fields associated with different domains in the ferroelectric layer~\cite{6}. Figure~\ref{FIG:6} shows intermediate ON resistance states measured for different set voltages (the pulse width for each set operation was changed to keep the same sweep rate). These set pulses with different voltage magnitudes can be considered as ‘partial-set’ operations. A full reset pulse of -4.5 V (10~ms pulse width) precedes each ‘partial-set’ operation. So before every ‘partial-set’ operation, it is considered that all the polarization domains are pointing away from Al$_{2}$O$_{3}$ layer leading to an OFF state (see Fig.~\ref{FIG:1}). 
As observed in Fig.~\ref{FIG:6}, different ON resistance values are obtained for ‘partial-set’ operations with pulse magnitude between 3 and 4.5~V. This corresponds to the coercive voltage range in the P-V curve for the stack shown in Fig.~\ref{FIG:4}. Therefore, one can consider that during each ‘partial-set’ operation, progressively more domains are switched leading to a gradual increase of the ON current with increasing magnitude of the ‘partial-set’ voltage. Thus, coercive voltage distribution of the domains can be utilized to achieve multiple resistance states in these FTJ stacks.

\begin{figure}
	\centering
	\includegraphics[scale=.15]{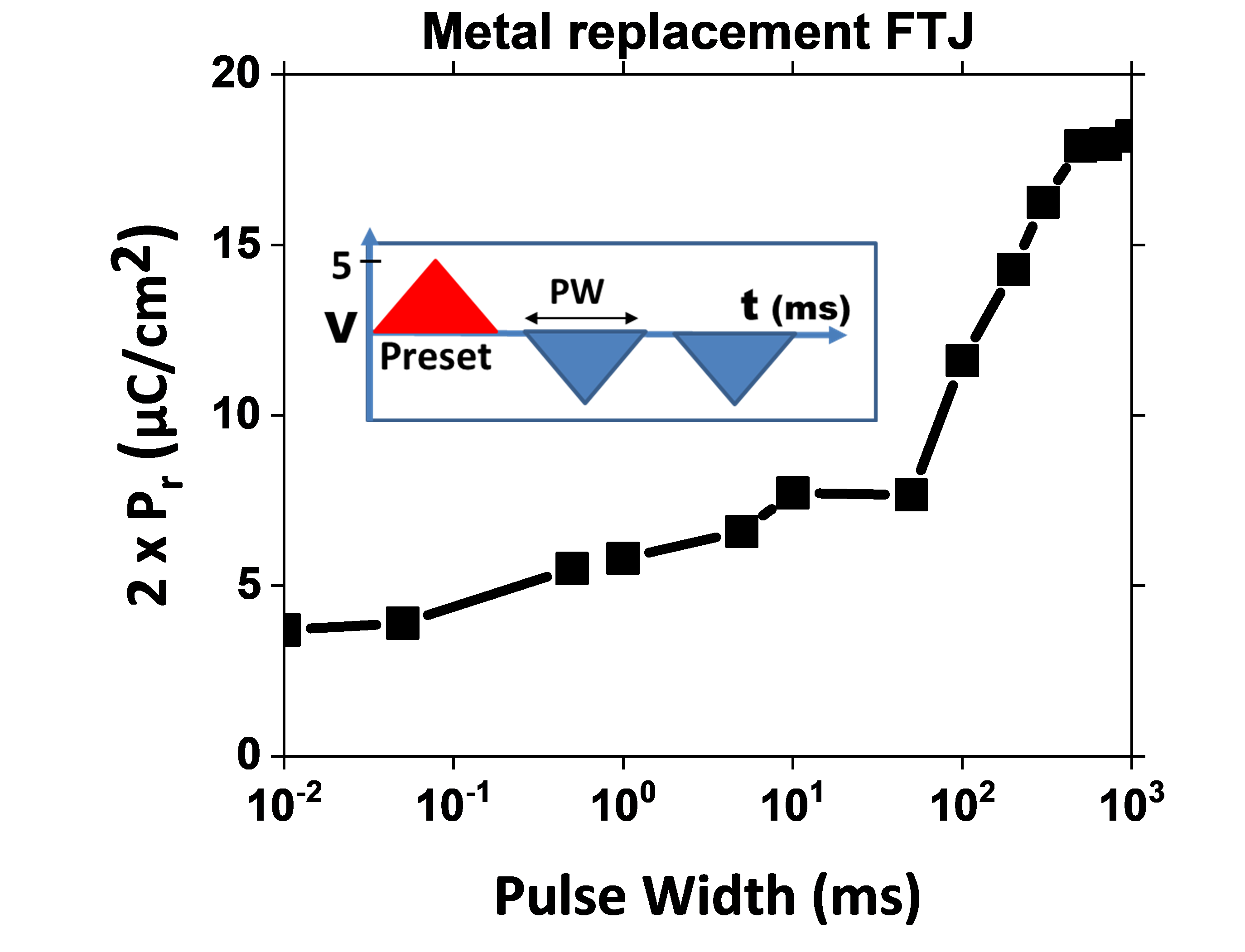}
	\caption{Polarization in the HZO layer estimated by ‘negative-down’ pulse sequence for various pulse widths in ‘metal replacement’ FTJ stack. The pulse sequence is shown in inset. ‘PW’ stands for ‘pulse width’.The preset voltage is +5~V and voltage for ‘negative-down’ pulses is -5~V.}
	\label{FIG:7}
\end{figure}

In order to understand the stabilization of the polarization in the HZO layer, the remnant polarization of the stack was estimated by measuring the polarization in ‘negative-down’ pulse sequences for various pulse widths (each measurement was preceded by a positive preset pulse). The measured polarization in the ‘metal replacement’ FTJ stack is shown in Fig.~\ref{FIG:7}. The inset shows the pulse sequence used.

As observed in Fig.~\ref{FIG:7}, the remnant polarization is found to increase with increasing pulse width. This implies that a higher polarization is stabilized in the stack for longer pulse widths. Three different behaviors are observed as a function of the pulse width. First, the increase in remnant polarization is gradual when the pulse width increases from 0.01 to 100 ms. Then the polarization value (2P$_{r}$) increases by a much larger magnitude for pulse widths in the range of 50 to 500 ms with a jump from 7.7~µC/cm$^{2}$ to 17.9~µC/cm$^{2}$ respectively. Finally, the polarization levels off ($\sim$18~µC/cm$^{2}$) for larger pulse widths. This behavior indicates that the stabilization of the ferroelectric polarization involves charging or discharging of traps near the ferroelectric-dielectric interface~\cite{7}. Such a mechanism requires sufficient time for the charge carriers to move to/from the trap states. A higher remnant polarization should in-turn lead to a higher TER in the stack. In this case, the polarity of the pulses corresponds to reset pulses and, hence, a lower OFF state current is observed with increasing pulse width (not shown). This measurement highlights the importance of charge traps in the operation of the Metal-Ferroelectric-Dielectric-Metal FTJ stack. Therefore, interface design with appropriate trap density is crucial to obtain high TER in such FTJ stacks.

\section{Conclusion}
CMOS BEOL compatible ferroelectric tunnel junction devices with 400°C HZO crystallization and W-HZO-Al$_{2}$O$_3$-TiN stack are demonstrated. This stack shows performance comparable to earlier works with higher temperature annealing and top TiN electrode. A ‘metal replacement’ process showing ON current density improvement compared to W top metal annealed stack is also reported. Partial switching operations are characterized, which allows multiple resistance states necessary for neuromorphic computing. Furthermore, a strong dependence of the remnant polarization magnitude with the pulse width is evidenced and reveals the importance of charge traps in providing sufficient charge screening for the stabilization of the polarization. Engineering the charge traps at the interface of the dielectric and the ferroelectric layer provides a key pathway to improve the design of FTJ device stacks.

\section{Acknowledgement}
Funding by Horizon 2020 EU project BeFerroSynaptic (No 871737) is acknowledged.


\end{document}